\documentclass[preprint,aps,superscriptaddress,nofootinbib]{revtex4}
\usepackage{graphicx}
\usepackage{amsbsy}

\begin{document}
\title{Vector Component of Confinement Force and\\ P-Wave Spectrum of $D$ mesons} 
\author{Jun Sugiyama} 
\email{sugiyama@th.phys.titech.ac.jp}
\affiliation{Department of Physics, Tokyo Institute of Technology, Tokyo 152-8551, Japan}
\author{Satoshi Mashita}
\affiliation{Department of Physics, Tokyo Institute of Technology, Tokyo 152-8551, Japan}
\author{Muneyuki Ishida}
\affiliation{Department of Physics, Meisei University, Hino 191-8506, Japan}
\author{Makoto Oka}
\affiliation{Department of Physics, Tokyo Institute of Technology, Tokyo 152-8551, Japan}

\begin{abstract}
The mass spectrum and the pionic decay widths of the P-wave $D$ mesons are studied using the Dirac equation for the light quark.
Our aim is to determine the Lorentz property of the confinement force.
We find that the Lorentz scalar confinement is consistent with the mass spectrum,
while a significant mixture of the Lorentz vector confinement is necessary to explain the decay widths.
\end{abstract}
\maketitle
\section{Introduction}

Confinement of color is the most outstanding problem in hadron physics.
Quantum chromodynamics (QCD) has revealed that the ground state of
pure gluonic QCD is in the confined phase, and that static (heavy) quarks
are confined as is shown by the area law behavior of Wilson loops.\cite{Bali}
Yet, confinement of dynamical quarks is not well established, although 
we believe that light quarks are confined as well and all hadrons are
white.

In phenomenology, on the other hand, various well-tuned quark models
are available in explaining hadron spectra.  The simplest system may be 
the heavy quarkonia, such as $\bar c c$ ($J/\psi$, $\psi'$ \ldots)
and $\bar b b$ ($\Upsilon$).
Their spectra are so simple as to be reproduced by a nonrelativistic
Hamiltonian quark model.\cite{Mukherjee} The Cornell potential,\cite{Eichten} which consists of
a linear confinement potential and a short-range color-Coulomb force,
reproduces overall level structure and also agrees very well with 
the lattice QCD static potential.  
The recoil (or relativistic) corrections, up to the order $1/M_Q^2$, 
can be computed for the single gluon exchange process and are shown 
to give reasonable spin dependences for fine structures of the spectra.\cite{DGG}

One important question, which has not been properly answered in the 
lattice QCD, is the covariance property of the confinement.
In the Hamiltonian quark model, as an approximation of the full QCD
dynamics, the inter-quark potential may have various Dirac structures
at the vertex.  For instance, the one-gluon exchange potential is
assumed to have a $\gamma^{\mu}$ at each vertex.  In the static
potential limit only the time component $\gamma^0$ survives, which
of course give a Coulomb potential.  On the other hand, if we exchange
a few gluons between the quarks, for instance, the Dirac structure 
at the vertex can be further complicated.  

In addressing the above problem, we choose a semi-simple system, 
that is, bound states of a heavy antiquark ($\bar Q$) and a light quark ($q$).
Concretely, we treat here the D-meson spectrum, which contains a
charm (anti)quark and a light quark.
This heavy-light system has a simple limit if the heavy quark mass $M_Q$ is infinite.  The system reduces to a single quark moving 
under a potential which is independent from the heavy quark spin.
The spin independence is a result of the heavy quark symmetry of QCD.\cite{IsgurW}
Now the light quark is supposed to move relativistically and therefore
we need to treat this system (in the large $M_Q$ limit) in terms of
the Dirac equation.\cite{Kaburagi}

Such Dirac equation is required to have a confinement potential, but
the Lorentz property of the confinement potential is not a priori given.
Without derivative (or velocity dependence), we have two choices 
in general.  Either a Lorentz scalar potential, which appears in the
Dirac Hamiltonian as $\beta S(r)$, or the time component of 
a Lorentz vector potential, given as $V(r)$.
Namely, the Dirac Hamiltonian of the form
\begin{equation}
H = \vec\alpha\cdot\vec p + \beta m + \beta S(r) +V(r)
\label{eq:Dirac-Hamiltonian}
\end{equation}
is supposed to represent the system.
Besides, we also need the color-Coulomb part of the interaction for
this system as well, which is attributed to the one-gluon exchange
interaction and is therefore included in the Lorentz vector part $V(r)$.

The purpose of the present paper is to study how we can determine 
the Lorentz property of the confinement by using the $\bar Q q$ 
meson observables, i.e., the masses and decay amplitudes.
For this purpose we employ a linear confinement potential 
whose total strength is fixed but the ratio of the Lorentz scalar
and vector components is allowed to vary,
\begin{equation}
V_{\rm conf} =  \beta (1-v) br + v br 
\label{eq:V-conf}
\end{equation}
The parameter $v$ denotes how much of the confinement is regarded as
the Lorentz vector.
By varying $v$, we study the D meson spectrum and decays and
attempt to determine which $v$ is the most appropriate.
We are particularly interested in the P-wave spectrum of the $D$ mesons,
because there the spin-orbit interaction plays an essential role.
It is well known that the Coulombic potential in the Dirac equation gives 
a splitting of the $p_{1/2}$ and $p_{3/2}$ states, where the $p_{1/2}$
is lower in energy than $p_{3/2}$.  The order of the two $p$ states is
determined by the sign of the effective spin-orbit potential, and  depends
on whether the potential is regarded as the Lorentz vector or scalar.\cite{Isgur}
Indeed this splitting is reversed when the Dirac equation for a pure scalar attractive potential
is solved regardless of the shape of the potential. 
The MIT bag model\cite{Chodos} is an example, where the ground $1s_{1/2}$ is followed by 
the 1st excited state $2p_{3/2}$ instead of $2p_{1/2}$.
It is therefore extremely interesting to find the order of the $2p_{1/2}$ and $2p_{3/2}$
states in the $D$ meson spectrum.

In reality, however, as the heavy quark spin is not completely decoupled, we need to take
into account $1/M_Q$ correction terms and especially the hyperfine splitting
due to the spin-spin interaction.

In Sect. \ref{ch:wave}, we present the Dirac equation and its solutions relevant to the discussion
in this paper.  In Sect. \ref{ch:1/MQcorrect}, $1/M_Q$ correction terms are introduced both for the 
scalar potential as well as the vector potential.
In Sect. \ref{ch:decay}, we calculate the pionic decay widths of the $D$ mesons.   The decay 
widths are computed in the first-order perturbation theory and agree with the
results from the heavy quark symmetry.
In Sect. \ref{ch:parameter}, we first explain how we determine the values of the parameters and 
then present the results of our calculation.
In Sect. \ref{ch:conclusion}, a conclusion and discussion are given.

\section{Wave Functions}\label{ch:wave}
We start with the simple picture of the $D$ mesons, i.e.,
a light quark bound in a potential created by a heavy anti-quark, which sits still at the origin.
We employ the Dirac equation with a linear plus Coulomb potential to describe the motion
of the light quark.\cite{Kaburagi}

As shown in Eq. (\ref{eq:V-conf}), we assume that 
the linear potential may include scalar and vector components. 
 \begin{equation}
 \mathcal{L}=\bar{\Psi}\left[i\partial\hspace{-0.55em}/-m+\beta\frac{a}{r}-br\{(1-v)+\beta v\}\right]\Psi
 \end{equation}
where $m$ is the constituent mass of light quark, $a=\frac{4}{3}\alpha_s$, and $b$ is the string tension.
The corresponding Hamiltonian is given by
 \begin{equation}
  \mathcal{H}=\mathbf{\boldsymbol{\alpha}}\cdot \mathbf{p}+\beta m-\frac{a}{r}+\beta(1-v)br+vbr \label{ham}
 \end{equation}

\begin{figure}[hbt]
  \centerline{\includegraphics{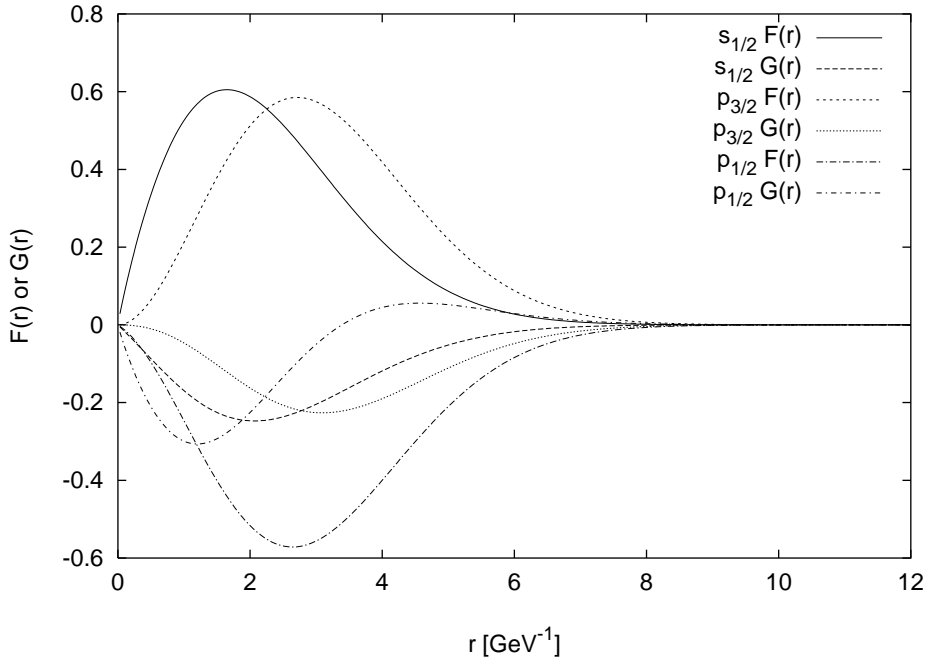}}
  \caption{Wave functions of $1s_{\frac{1}{2}}$, $2p_{\frac{3}{2}}$ and $2p_{\frac{1}{2}}$ states. ($m\! =\! 300\mathrm{MeV}\quad\frac{4}{3}\alpha_s\! =\! 0.360\quad b\! =\! 0.190\mathrm{GeV}^2 \quad v\! =\! 0\%$)}
  \label{fig:wf}
 \end{figure}

The total angular momentum $\vec j$ is conserved for the central potential in Eq. (\ref{ham}), and then
the angular part of the wave function is written by the spinor spherical harmonics,
 \begin{equation}
  \mathcal{Y}^{\ell}_{jm}(\theta,\phi)\equiv\sum_{m_\ell,m_s}\langle \ell\> m_\ell \> s \> m_s | j \> m\rangle Y_{\ell m_\ell}(\theta,\phi)\chi_{m_s}
 \end{equation}
We parametrize the wave function as
 \begin{equation}
  \Psi^{\ell=j\pm \frac{1}{2}}_{jm} (r,\theta,\phi)=\frac{1}{r}\left(
   \begin{array}{c}
    F(r)\mathcal{Y}^{\ell=j\pm \frac{1}{2}}_{jm}\\
    iG(r)\mathcal{Y}^{\ell^\prime=j\mp \frac{1}{2}}_{jm}
   \end{array}\right) \label{wave}
 \end{equation}
where $\ell$ and $\ell^\prime$ denote the orbital angular momentum of the large and small components respectively.
Then the equations of motion for the radial parts are given by
@\begin{eqnarray}
    \left(m+(1-v)br+vbr-\frac{a}{r}-E\right)F(r)&=&\left(\frac{d}{dr}-\frac{\kappa}{r}\right)G(r)\nonumber\\
    \left(-m-(1-v)br+vbr-\frac{a}{r}-E\right)G(r)&=&\left(-\frac{d}{dr}-\frac{\kappa}{r}\right)F(r) \label{dirac}
 \end{eqnarray}
where $\kappa=\pm(j+\frac{1}{2})$ for $j=\pm\frac{1}{2}$. Explicitly, $\kappa=-1$ for $1s_{\frac{1}{2}}$, $\kappa=-2$ for $2p_{\frac{3}{2}}$ and $\kappa=1$ for $2p_{\frac{1}{2}}$. 
$E$ is the energy eigenvalue, for which we only consider the positive energy solutions.
Wave functions and energy eigenvalues are acquired by solving this equation.
Some of the obtained wave functions are shown in Fig.~\ref{fig:wf}.
We will use them to calculate the $1/M_Q$ correction to the mass spectrum as well as the decay widths in the following sections.

\section{$1/M_Q$ correction terms for the $D$ meson masses}\label{ch:1/MQcorrect}

Once we solve the Dirac equation for the light quark in the linear plus Coulomb potential, 
we are ready to compute the mass spectrum of the $D$ mesons.  
In the heavy quark limit, the heavy quark spin is decoupled and we have sets of degenerate states.
Among them, here we are interested in 
$D$ ($J^\pi=0^-$) and $D^*$ ($J^\pi=1^-$) states from the $1s_{\frac{1}{2}}$ eigenstate of the light quark,
$D^*_0$ ($J^\pi=0^+$) and $D^*_1$ ($J^\pi=1^+$) from $2p_{\frac{1}{2}}$, and 
$D_1$ ($J^\pi=1^+$) and $D^*_2$ ($J^\pi=2^+$) from $2p_{\frac{3}{2}}$.
In the real world, the charm quark is not infinitely heavy so that we have splittings in each
pair. The most popular one is the spin-spin splitting in the S states, or $D-D^*$ splitting.
The splittings are commonly attributed to the interaction of higher orders in $1/M_Q$ expansion, where $M_Q$ is 
the heavy quark mass.  
It is also important to note that the $D_1$ and $D^*_1$ states may mix with each other when the heavy
quark spin is allowed to flip.  Thus, the $1/M_Q$ higher order terms will mix  $D_1$ and $D^*_1$.

We here derive the $1/M_Q$ interaction terms associated with the linear plus Coulomb potential,
which are regarded as relativistic corrections coming from the Lorentz scalar and/or vector couplings.
The corrections for the Coulomb part is similar to the Fermi-Breit interaction for the heavy $Q-\bar Q$
system.  We consider the one-gluon exchange diagram in the Coulomb gauge,
which gives the gluon propagator
 \begin{eqnarray}
  D_{00}&=&-\frac{4\pi}{\mathbf{q}^2}\nonumber\\
  D_{ik}&=&\frac{4\pi}{\mathbf{q}^2-{q_0}^2}\left(\delta_{ik}-\frac{\mathbf{q}_i\mathbf{q}_k}{\mathbf{q}^2}\right)
  \cong\frac{4\pi}{\mathbf{q}^2}\left(\delta_{ik}-\frac{\mathbf{q}_i\mathbf{q}_k}{\mathbf{q}^2}\right)+\mathcal{O}\left(\frac{1}{M_Q}\right) 
 \end{eqnarray}
 and compute the $q\bar Q$ scattering (invariant) amplitude to the leading order,
 \begin{eqnarray}
  \mathcal{M}&=&-\frac{4\alpha_s}{3}({\bar{u}_1}^\prime\gamma^\mu u_1) D_{\mu\nu}(q)({\bar{u}_2}^\prime\gamma^\nu u_2)\nonumber\\
                  &=&-\frac{4\alpha_s}{3}\left\{({\bar{u}_1}^\prime\gamma^0 u_1)({\bar{u}_2}^\prime\gamma^0 u_2)D_{00}+
                      ({\bar{u}_1}^\prime\gamma^i u_1)({\bar{u}_2}^\prime\gamma^k u_2)D_{ik}\right\} \label{ampli}
 \end{eqnarray}
Here, $u_1$ and $u_2$ denote the plane wave Dirac spinors of the light and heavy quarks, respectively.
Then, we replace the heavy quark spinors in  (\ref{ampli}) by their static expansion, 
 \begin{equation}
  u_2(\mathbf{p})=\frac{1}{\sqrt{2E_Q}}\left(
   \begin{array}{c}
    \sqrt{E_Q+M_Q}\> \chi\\
    \sqrt{E_Q-M_Q} \mathbf{n}\cdot\boldsymbol{\sigma}\>\chi
   \end{array}
  \right)
 \end{equation}
and the amplitude up to the order to $1/M_Q$ is given by
 \begin{equation}
  \mathcal{M}=-a\biggl\{{(\bar{u}_1}^\prime\gamma^0 u_1)\!\!\underbrace{{\chi_2^\prime}^\dagger \chi_2}_{\textrm{1st part}}\!\! D_{00}
  +\frac{1}{M_Q}({\bar{u}_1}^\prime\gamma^i u_1){\chi_2^\prime}^\dagger\Bigl(\underbrace{i\boldsymbol{\sigma}\times(-\mathbf{q})}_{\textrm{2nd part}}
  +\underbrace{2\mathbf{p}_2-\mathbf{q}}_{\textrm{3rd part}}\Bigr)^k \chi_2 D_{ik}\biggr\}
 \end{equation}
 
 Now, we determine the effective potential (which is in general, velocity dependent)
 so as to recover this amplitude in the Born approximation. 
The potential corresponding to the first part is the Coulomb potential, $-a/r$, which is of order $M_Q^0$.
Energy eigenvalue corresponding to this is already acquired by solving (\ref{dirac}).
The second and third parts induce $1/M_Q$ terms. 
The potential corresponding to the second part is given by
 \begin{equation}
  \frac{a}{M_Q}\boldsymbol{\sigma}_1\cdot\left(
   \begin{array}{cc}
    0&\displaystyle \mathbf{S}_Q\times \frac{\mathbf{r}}{r^3} \\
    \displaystyle \mathbf{S}_Q\times \frac{\mathbf{r}}{r^3} & 0
   \end{array}\right) \label{eq:12}
 \end{equation}
Because this term may change the spin of the heavy quark, this leads to a mixing between $D_1$ and $D_1^*$.
The potential corresponding to the third part is given by
 \begin{equation}
  -\frac{a}{M_Q}\frac{1}{2r}\left(
   \begin{array}{cc}
    0&\displaystyle \boldsymbol{\sigma}\cdot\mathbf{p}+ \frac{\boldsymbol{\sigma}\cdot\mathbf{r}}{r^2}\mathbf{r}\cdot\mathbf{p} \\
    \displaystyle \boldsymbol{\sigma}\cdot\mathbf{p}+ \frac{\boldsymbol{\sigma}\cdot\mathbf{r}}{r^2}\mathbf{r}\cdot\mathbf{p} & 0
   \end{array}\right) \label{eq:13}
 \end{equation}

 Similarly, the $1/M_Q$ terms generated by the linear potential can be obtained.  
 There we need to distinguish the Lorentz scalar part and the Lorentz vector part of the linear potential
 and calculate the contribution of each term separately.
 \begin{equation}
  \frac{bv}{M_Q}\boldsymbol{\sigma}_1\cdot\left(
   \begin{array}{cc}
    0&\displaystyle \mathbf{S}_Q\times \frac{\mathbf{r}}{r} \\
    \displaystyle \mathbf{S}_Q\times \frac{\mathbf{r}}{r} & 0
   \end{array}\right)+\frac{bv}{M_Q}\frac{1}{r}\left(
   \begin{array}{cc}
    0&\displaystyle r^2\boldsymbol{\sigma}\cdot\mathbf{p}+ \frac{\boldsymbol{\sigma}\cdot\mathbf{r}}{2i} \\
    \displaystyle r^2\boldsymbol{\sigma}\cdot\mathbf{p}+ \frac{\boldsymbol{\sigma}\cdot\mathbf{r}}{2i} & 0
   \end{array}\right) \label{eq:14}
 \end{equation}

Finally, one more $1/M_Q$ correction comes from the kinetic energy (recoil) of the heavy quark, i.e., 
 \begin{equation}
  \mathbf{p}^2/2M_Q \label{eq:15}
 \end{equation}
where $p$ is the momentum of the heavy quark.

The $1/M_Q$ Hamiltonians (\ref{eq:12}),(\ref{eq:13}),(\ref{eq:14}) and (\ref{eq:15}) are treated in the first order perturbation theory.
As a result we obtain the masses of $D^*$ and $D$ for the ground states, and those of $D_2$, $D_1$, $D_1^*$ and $D_0^*$ for the first excited states.
Here, we consider the mixing between two states $D_{\frac{1}{2}}$ and $D_{\frac{3}{2}}$,
and obtain $D_1^*$ and $D_1$ as physical states.
The detailed formulas are given in Appendix. \ref{ch:massformula}.

\section{Pionic Decay Widths of $D$ mesons}\label{ch:decay}
The pionic decay widths of the $D$ mesons are calculated by using PCAC relation.\cite{Weinberg} 
The decay widths into $D\pi$ states are given\cite{Yaouanc} by
 \begin{equation}
  \Gamma(D_n\to D \pi)=\frac{3}{8\pi f_{\pi}^2}\frac{|\mathbf{q}|^3}{2j_n+1}|X_{DD_n}|^2 \qquad(f_{\pi}=93\mathrm{MeV})
 \end{equation}
where $\mathbf{q}$ is the momentum of the emitted pion, and the amplitude $X_{DD_n}$ is given by
 \begin{equation}
  X_{DD_n}=\int d\mathbf{r} \> \Psi_{D}^\dagger(\mathbf{r})\left(\gamma_5-\frac{\mathbf{q}}{|\mathbf{q}|}\cdot\Sigma\right)\Psi_{D_n}(\mathbf{r})e^{-i\mathbf{q}\cdot\mathbf{r}}
 \end{equation}
Here we neglect the final state interaction.
Taking the direction of $\mathbf{q}$ along the $z$-axis, we obtain
 \begin{equation}
  X_{DD_n}=\sum_{\ell=0}^\infty(-i)^\ell\sqrt{(2\ell+1)(4\pi)}\int\! d\mathbf{r}\> \Psi_D^\dagger(\mathbf{r})(\gamma_5-\Sigma_3)\Psi_{D_n}\!(\mathbf{r})j_\ell(qr)Y^0_\ell(\mathbf{r})
 \end{equation}
From the conservation of parity and angular momentum, one can easily see that
$D^*$, $D_0^*$, and $D_2^*$ can decay into $D\pi$, while $D_1^*$ and $D_1$ cannot.
The explicit forms of $X_{DD_n}(D_n=D^*,D_0^*$ and $D_2^*)$ are given by 
 \begin{eqnarray}
  X_{DD^*}(q)\!\!&=&\!\!\!\int\left[\left\{-|F_0(r)|^2+\frac{1}{3}|G_0(r)|^2\right\}j_0(qr)+\frac{4}{3}|G_0(r)|^2j_2(qr)\right] dr\nonumber\\
  X_{DD_0^*}(q)\!\!&=&\!\!\!\int \biggl[i\!\left\{F_0(r)G_1(r)-G_0(r)F_1(r)\right\}j_0(qr)\nonumber\\
                    & &\hspace{2em}-i\!\left\{F_0(r)F_1(r)+G_0(r)G_1(r)\right\}j_2(qr)\biggr]dr \label{eq:decayampli}\\
  X_{DD_2^*}(q)\!\!&=&\!\!\!\int \left[\sqrt{2}i\!\left\{F_0(r)F_3(r)-\frac{1}{5}G_0(r)G_3(r)\right\}j_1(qr)\right.\nonumber\\
                     & &\hspace{2em}\left.+\sqrt{2}i\!\left\{F_0(r)G_3(r)-G_0(r)F_3(r)\right\}j_2(qr)
                          -\sqrt{2}i\left\{\frac{6}{5}G_0(r)G_3(r)\right\}j_3(qr)\right]dr\nonumber
 \end{eqnarray}
The radial wave functions for the $1s_{\frac{1}{2}}$, $2p_{\frac{1}{2}}$ and $2p_{\frac{3}{2}}$ states are labeled by $0$, $1$ and $3$, respectively.
For the decays into the $D^*\pi$ final channel, we use helicity amplitude $X_{D^*D_n}^{h_{D^*}h_{D_n}}$,
where the $h_{D^*}$ and $h_{D_n}$ denotes the z-component of the spin of initial $D_n$ (final $D^*$) meson. 
Since the z-component of the angular momentum is conserved, we obtain
 \begin{equation}
   \Gamma(D_n\to D^* \pi)=\frac{3}{8\pi f_{\pi}^2}\frac{|\mathbf{q}|^3}{2j_n+1}\left\{|{X}_{D^*D_n}^{++}|^2+|{X}_{D^*D_n}^{00}|^2+|{X}_{D^*D_n}^{--}|^2\right\}
 \end{equation}
Now, $D_2^*$, $D_1$ and $D_1^*$ can decay into $D^*\pi$, while $D_0^*$ cannot.
Using the matrix elements given in Eq. (\ref{eq:decayampli}), the decay width of $\Gamma(D_2^* \to D^* \pi)$ is given by
 \begin{equation}
  \Gamma(D_2^*\to D^* \pi)=\frac{3}{2}\frac{3}{8\pi {f_\pi}^2}\frac{|\mathbf{q}|^3}{5}|X_{DD_2^*}|^2 \label{eq:21}
 \end{equation}
We consider the mixing of $D_1$ and $D_1^*$ according to Eq. (\ref{mixmat}).
Denoting the eigenvectors of Eq. (\ref{mixmat}) by $(\alpha,-\beta)$ and $(\beta,\alpha)$, we obtain
 \begin{eqnarray}
  \Gamma(D_1 \to D^* \pi)(q)\!&=&\!\frac{3|\mathbf{q}|^3}{8\pi {f_\pi^2}}\left|\frac{1}{\sqrt{2}}\alpha X_{DD_2^*}(q)-\beta X_{DD_0^*}(q)\right|^2 \nonumber\\
  \Gamma(D_1^* \to D^* \pi)(q)\!&=&\!\frac{3|\mathbf{q}|^3}{8\pi {f_\pi^2}}\left|\frac{1}{\sqrt{2}}\beta X_{DD_2^*}(q)+\alpha X_{DD_0^*}(q)\right|^2 \label{eq:22}
 \end{eqnarray}

The heavy quark symmetry\cite{IsgurW} predicts the relations,
 \begin{eqnarray}
  \Gamma(D_2^*\to D \pi)\!&=&\!\frac{2}{5}\Gamma(D_{\frac{3}{2}} \to D^* \pi)\nonumber\\
  \Gamma(D_2^*\to D^* \pi)\!&=&\!\frac{3}{5}\Gamma(D_{\frac{3}{2}} \to D^* \pi)\label{heavysym}\\
  \Gamma(D_0^* \to D \pi)\!&=&\!\Gamma(D_{\frac{1}{2}}\to D^*\pi) \nonumber
 \end{eqnarray}
One easily sees from Eqs. (\ref{eq:21}) and (\ref{eq:22}) that these relations are satisfied if we use the same $|\mathbf q|$ for
all the decay widths.
In reality, due to the phase space differences, these relations are not exactly met. 

When we later compare calculations with experiment, 
we use  the observed total decay widths of the $D$ mesons,
because the pionic decay is dominant in the $D$ meson decays.
\section{Determination of Parameters and Results}\label{ch:parameter}
The Dirac equation has five free parameters;
the constituent light quark mass $m$, the strong coupling constant $\alpha_s$, the string tension $b$, the constituent heavy quark mass $M_Q$, and the zero point of energy ${m_c}^0$.
The ambiguity of the zero point corresponds to uncertainty of ${m_c}^0$ in (\ref{masss1}), (\ref{massp3}) and (\ref{massp1}).

These parameters are to be determined phenomenologically with some conditions imposed.
First, the string tension is predicted by the lattice calculation\cite{Bali}, so as to be $0.17\sim 0.20\mathrm{GeV^2}$.
The constituent charm quark mass should be about the half of the mass of $J/\Psi$, which is about $1.5\mathrm{GeV}$.
We will leave out unphysical region of parameters, which are too different from these values.

Now, these parameters are determined by using the well-determined masses of the $D$ meson family, which are $D$, $D^*$, $D_1$ and $D_2^*$.
Paying attention to the differences of the masses of the $D$ mesons instead of their absolute values removes ambiguity of the zero point. 
Then, four parameters remain. But, the differences of the masses of $D$ mesons give only three constraints.
So, we cannot fit all the parameters. 
Therefore, we choose the constituent light quark mass $m$ as $300$MeV.
The difference between $D$ and $D^*$, which is $141.45$MeV, determines $M_{Q}$ through $\langle \frac{1}{M_Q}(\frac{a}{r^2}+bv)\rangle$ in (\ref{masss1}).
After that, we determine the values of $a$ and $b$, by taking into account the experimental values of $D_1-D^*$ mass splitting ($413.85\mathrm{MeV}$) and of $D_2^*-D_1$ mass splitting ($36.8\mathrm{MeV}$) .
 \begin{figure}
  \begin{center}
   \includegraphics[width=8.5cm]{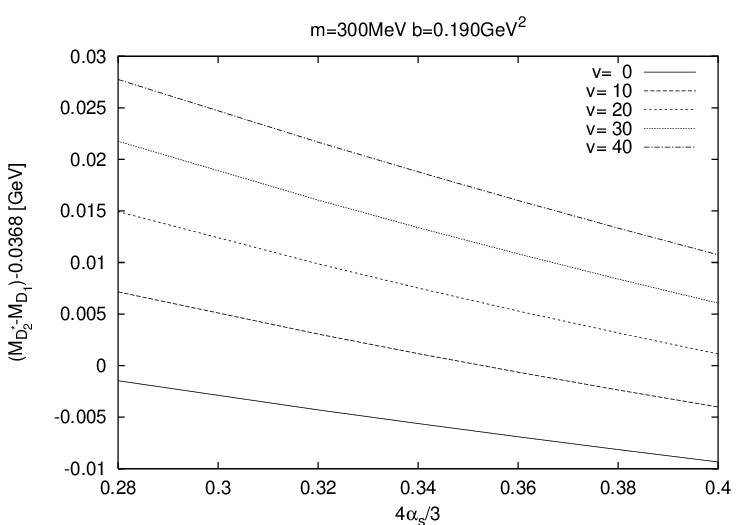}
   \caption{Difference between the experimental data of $D_2^*-D_1$ and our calculation
            for various $\alpha_s$ and $v$. }
   \label{fig:alphadif1}
   \includegraphics[width=8.5cm]{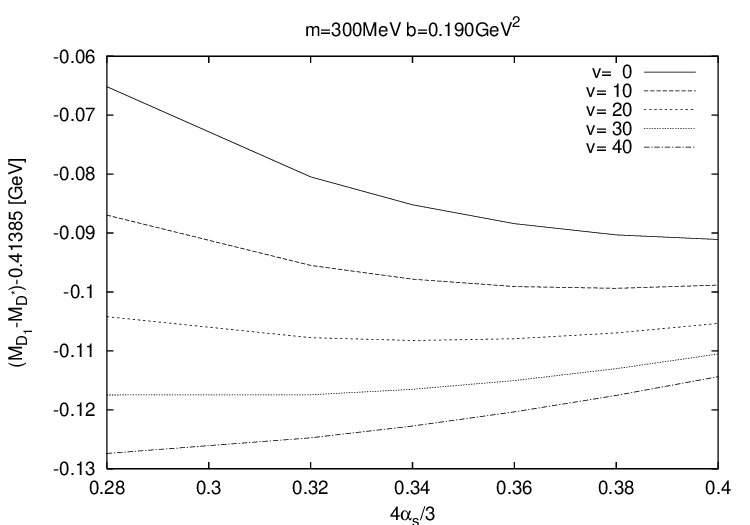}
   \caption{Difference between the experimental data of $D_1-D^*$ and our calculation
            for various $\alpha_s$ and $v$. }
   \label{fig:alphadif2}
   \includegraphics[width=8.5cm]{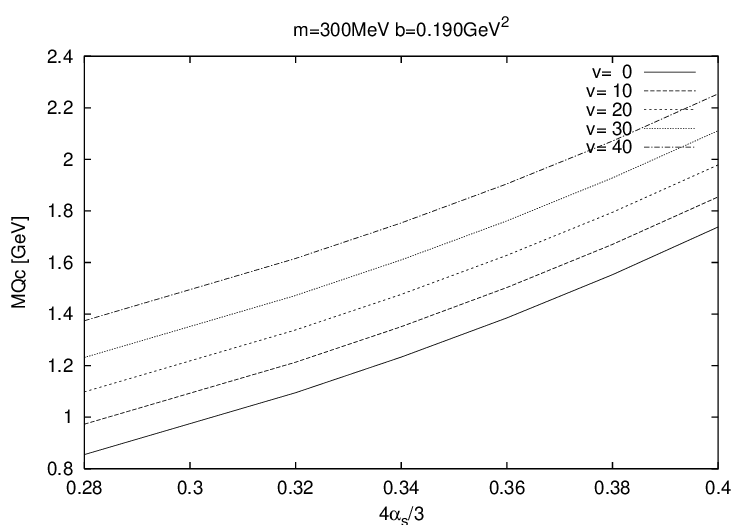}
   \caption{The constituent charm quark mass for various $\alpha_s$ and $v$. }
   \label{fig:alphadif3}
  \end{center}
 \end{figure}
 \begin{figure}
  \begin{center}
   \includegraphics[width=8.5cm]{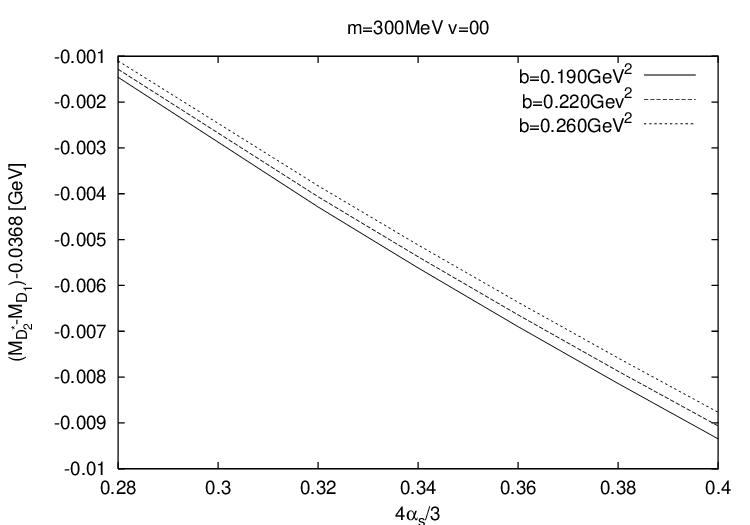}
   \caption{Difference between the experimental data of $D_2^*-D_1$ and our calculation
            for various $\alpha_s$ and $b$. }
   \label{fig:tension1}
   \includegraphics[width=8.5cm]{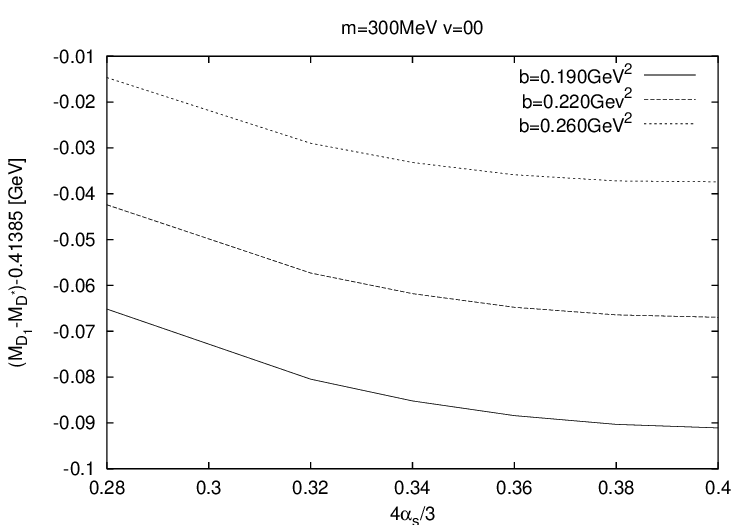}
   \caption{Difference between the experimental data of $D_1-D^*$ and our calculation
            for various $\alpha_s$ and $b$. }
   \label{fig:tension2}
   \includegraphics[width=8.5cm]{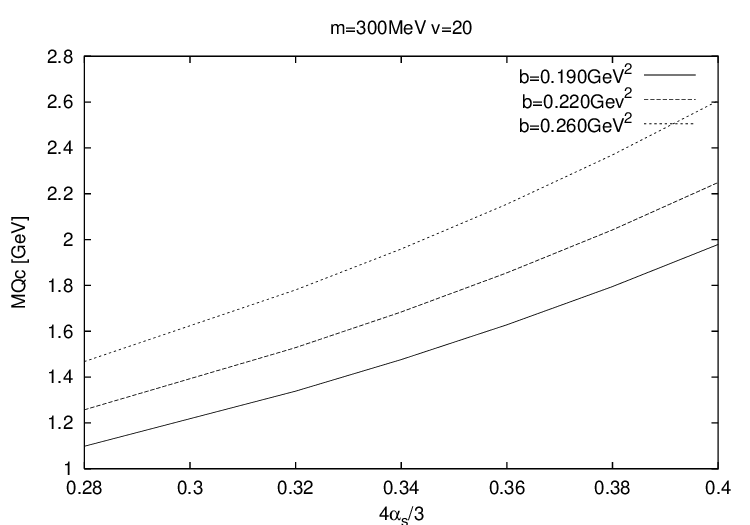}
   \caption{The constituent charm quark mass for various $\alpha_s$ and $b$.}
   \label{fig:tension3}
  \end{center}
 \end{figure}

The parameter dependences of various quantities are shown in Figs.~\ref{fig:alphadif1}-\ref{fig:tension3}.
From Fig.~\ref{fig:tension2}, one sees that 
in order to reproduce the $D_1-D^*$ mass splitting we must use a large value of $b$ about $0.26\mathrm{GeV}^2$.
However, this value is not acceptable from the lattice calculation.
Furthermore, in the region of $\frac{4}{3}\alpha_s>0.35$ by taking such a large value of $b$,
we are forced to use too large value of $M_{Q_c}$ in order to reproduce the experimental value of $D^*-D$ mass splitting. (See Fig.~\ref{fig:tension3}.)

Thus we give up reproducing the $D_1-D^*$ mass splitting, and employ $b = 0.19\, \mathrm{GeV}^2$, which is
within the lattice prediction.
In this case the predicted $D_1-D^*$ mass splitting is about $340\mathrm{MeV}$, which is $70\mathrm{MeV}$ smaller than the experimental value.

From Fig.~\ref{fig:alphadif1} one notices that small (large) $\alpha_s$ is favored for small (large) $v$.
However, because of the constraint on the value of $M_{Q_C}$ shown in Figs.~\ref{fig:alphadif3} and \ref{fig:tension3}, 
we cannot let $\alpha_s$ too small (large).
We therefore choose $a=\frac{4}{3}\alpha_s$ within the region $0.32 \leq a(={\frac{4}{3}\alpha_s}) \leq 0.40$
in order to reproduce the experimental $D^*-D$ mass-splitting. 
The value of ${m_c}^0$ is fixed by $m(D_2^*)=2459\mathrm{MeV}$, which is the most reliable among the P-wave states.
 \begin{figure}[tb]
 \begin{center}
  \includegraphics[width=8.5cm]{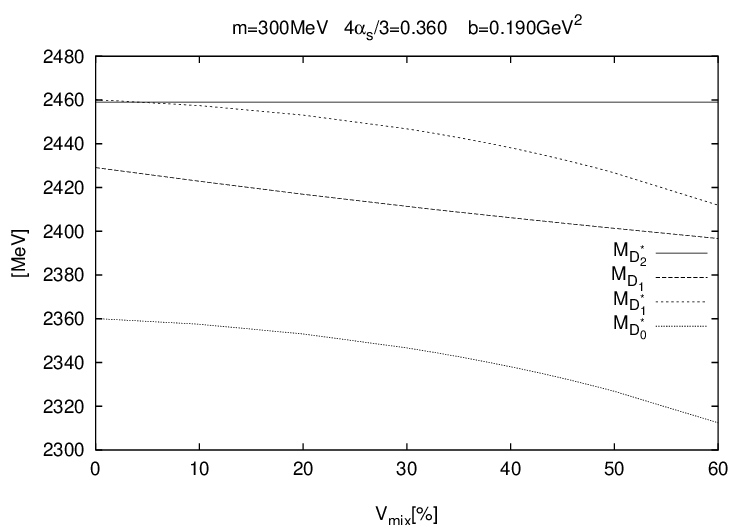}
  \caption{The masses of $D$ mesons in P-wave. We fit $m(D_2^*)=2459\mathrm{MeV}$.
           The value of parameters are $\frac{4}{3}\alpha_s=0.36, \quad b=0.19\mathrm{GeV^2}$ and $m=300\mathrm{MeV}$.}
  \label{fig:d2fit}
  \includegraphics[width=8.5cm]{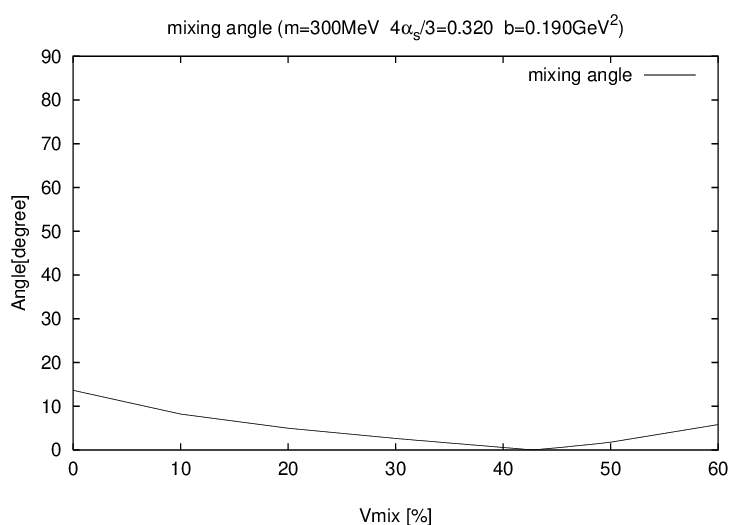}
  \caption{The mixing angle of $D_1$ and $D_1^*$.
           The value of parameters are $\frac{4}{3}\alpha_s=0.32, \quad b=0.19\mathrm{GeV^2}$ and $m=300\mathrm{MeV}$.}
  \label{fig:angle}
 \end{center}
 \end{figure}
 \begin{figure}
 \begin{center}
  \includegraphics[width=8.5cm]{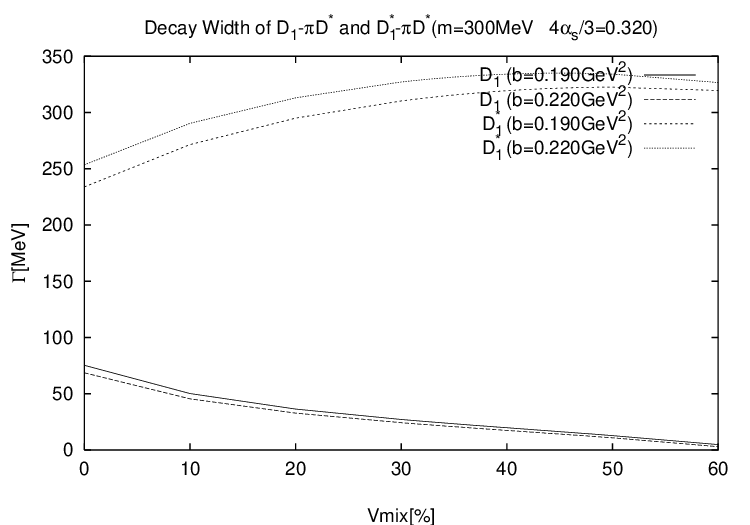}
  \caption{The pionic decay widths of $D_1$ and $D_1^*$. 
           The value of parameters are $\frac{4}{3}\alpha_s=0.32, \quad b=0.19\mathrm{GeV^2}$ and $m=300\mathrm{MeV}$.}
  \label{fig:decayd1}
   \includegraphics[width=8.5cm]{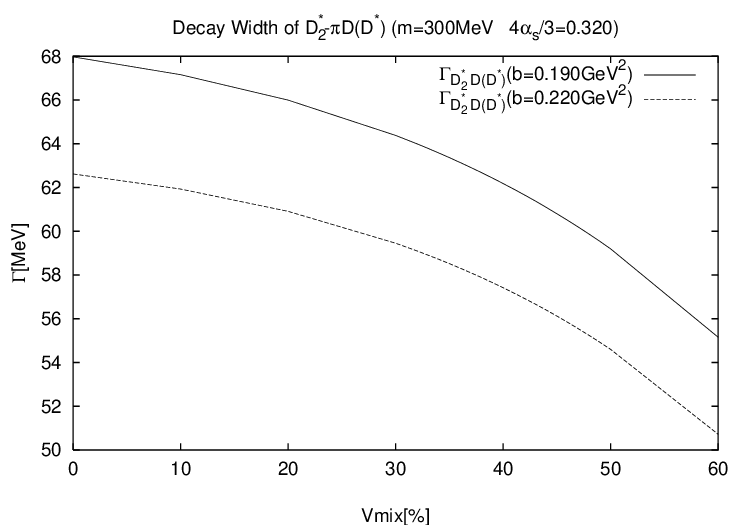}
  \caption{The pionic decay width of $D_2^*$.
           The value of parameters are $\frac{4}{3}\alpha_s=0.32, \quad b=0.19\mathrm{GeV^2}$ and $m=300\mathrm{MeV}$.}
  \label{fig:decayd2st}
  \includegraphics[width=8.5cm]{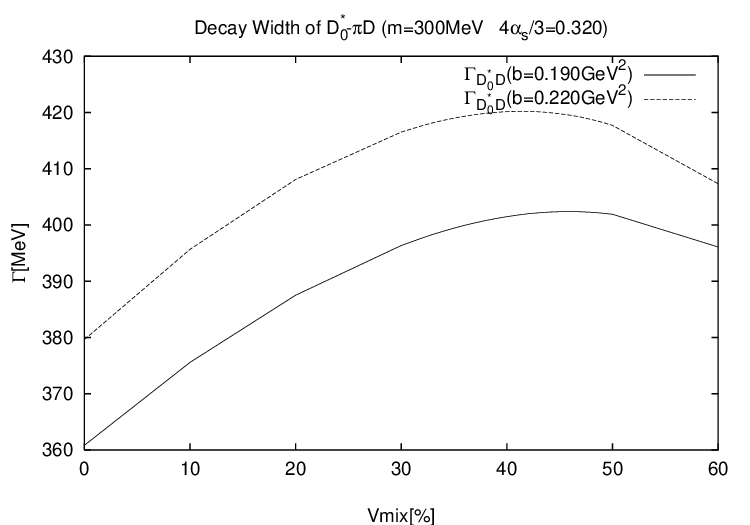}
  \caption{The pionic decay width of $D_0^*$.
           The value of parameters are $\frac{4}{3}\alpha_s=0.32, \quad b=0.19\mathrm{GeV^2}$ and $m=300\mathrm{MeV}$.}
  \label{fig:decayd0st}
 \end{center}

 \end{figure}

The results are shown in Fig.~\ref{fig:d2fit}.
This figure indicates that $D_1^*$ does not cross $D_1$ in the case of $0\%\leq v \leq 60\%$.
From Fig.~\ref{fig:angle}, one sees that the mixing angle of $D_1$ and $D_1^*$ decreases as $v$ increases,
while the difference between the eigenvalues of $M(D_{\frac{3}{2}})$ and $M(D_\frac{1}{2})$ decreases as $v$ increases.
As the mixing angle is small,
$D_1^*$ is dominated by $D_{\frac{1}{2}}$, and $D_1$ is dominated by $D_{\frac{3}{2}}$ 
in all region of $v$.
This is caused by the off diagonal elements of the mixing matrix given by (\ref{mixoff}).
As is seen from Fig.~ 1, the $F_3(r)G_1(r)$ product in Eq.~(\ref{mixoff}) changes its sign at around
$r\sim 3.0\,\mathrm{GeV^{-1}}$.
Because the Coulomb term is dominant at short distances and the linear potential is dominant 
at large distances, they tend to cancel when $v$ increases.

The pionic decay widths of $D_1$ and $D_1^*$ are more sensitive to the mixing effect than their masses,
because the difference between $\Gamma(D_{\frac{3}{2}}\to D^*)$ and $\Gamma(D_{\frac{1}{2}}\to D^*)$ is very large.
The $D_{\frac{1}{2}}$($D_{\frac{3}{2}}$) decaying into $D^*\pi$ occurs mainly in the S-wave (D-wave), so the decay width is large (small).
Then the difference of the decay widths $\Gamma(D_1 \to D^*)$ and $\Gamma(D_1^* \to D^*)$ decreases as the mixing angle is increased.

We compare our results with experimental data from Belle\cite{Belle}, CLEO\cite{CLEO}, and Particle Data Group\cite{PDG}.
$M(D_1^*)$ is larger than $M(D_1)$ in the CLEO data, while the order is reversed in the Belle data.
Our results agree with the former. 
In the case of small $v$, the masses of the P-wave $D$ mesons are consistent with the data of CLEO.
But, the difference of the decay widths $\Gamma(D_1 \to D^*)$ and $\Gamma(D_1^* \to D^*)$ is smaller than the observed value.
This problem is settled by increasing $v$ according to the above argument. 
Choosing $v=40\%$, we can reproduce $\Gamma(D^*_1 \to D^*)/\Gamma(D_1 \to D)$ rather well.
But, in this case, P-wave spectrum does not reproduce the experimental data.
Furthermore, the charm quark mass is required to be heavy, $\sim 2\mathrm{GeV}$.

The results are summarized as follows.
\vspace{-0.2em}
 \begin{enumerate}
  \item The mass spectrum of the P-wave $D$ mesons indicates that the confinement force is almost scalar or, even if vector component exists, it is about 10 percent at most.
  \item On the other hand, the pionic decay widths indicate that the confinement force includes significant amount of vector component.
 \end{enumerate}
\section{Conclusion and Discussion}\label{ch:conclusion}

The heavy-light quarkonium spectra are suitable for the study of the Lorentz property
of the color confinement force.  Recent new data of the $P$ wave spectrum of
the $D$ meson  systems are quite important and powerful in determining the spin-dependent
interactions among quarks.
We concentrate on the difference of the spin-orbit splitting from the Lorentz scalar and vector
quark confinement, and attempt to determine how much Lorentz vector confinement is necessary
to account for the $P$ wave spectrum of the $D$ mesons.

For this purpose, we start with the heavy quark limit in which the motion of the light quark 
is treated by a single particle Dirac equation. The light quark is bound by a potential that
consists of the confinement and one-gluon-exchange pieces.  By varying the vector
component of the confinement, we study the spectrum as well as the pionic decay widths of
the $D$ mesons.

We pick up three choices of $v$;  A: $v=0$ (pure scalar confinement), B: $v=0.1$, and
C: $v=0.4$.  The results are summarized in Table.~\ref{ta:results}.
A problem common to all the parameter choices is that the splitting between $1S$ and $2P$
states is too small.  We have found that unless we change some unnaturally large confinement
potential, we are not able to reproduce the difference.  Thus we give up fitting the parameters
for this splitting.

When we tune up the parameters mostly to the $P$-wave spectrum, then we obtain set A and 
set B, while the decay widths happen to prefer set C.
The set A and B are consistent with the mass spectrum obtained at CLEO, but the pionic decay widths are
not satisfactory, though the set B is slightly better.
We note, however, that the recent data from Belle group predict the $D_1^*$ mass
much smaller than that given by CLEO, and in fact the order of $D_1^*$ and $D_1$ 
is reversed.
Within our searches of the parameters, we never see the reversal of this $J=1$ states.
It is highly desirable to determine the mass of $D_1^*$ precisely in order to reduce the
vector amount of the confinement.

The set C is unique as it predicts significant amount of Lorentz vector component in the confinement.
The results are consistent with the pionic decay widths, but show a worse fit to the mass spectrum, 
and also the charm quark mass becomes large.
As the experimental data for the decay widths of $D_1$ and $D_2^*$ are in good quality, it seems
preferable to use these data to fix the vector amount of the confinement.  We then see that
a larger vector confinement gives a better fit.
At the same time we find that the mixing of $p_{1/2}$ and $p_{3/2}$ states in the $J=1$ mesons,
$D_1$ and $D_1^*$, decreases when the vector part of the confinement increases.  This causes 
the pionic decay width of $D_1$ to get lower, as the pure $p_{3/2}$ state cannot decay into
$\pi$ and $S$ state by keeping the conserving the angular momentum and parity simultaneously.
Thus the experimental data of the pionic decays suggest that the mixing is suppressed.
In our approach this leads to the conclusion that the vector component of the confinement force
must be significant in order to reduce the mixing strength.

Similar approaches have been taken by Ebert {\it et al}.\cite{Ebert} and also by Di Pierro {\it et al}.\cite{DiPierro}.
The former considered the $\bar{Q}q$ systems in terms of the Dirac equation with $1/M_Q$ corrections.
They considered mixing of Lorentz vector confinement as well as the anomalous color-magnetic coupling.
They fixed the ratio of the Lorentz vector and scalar confinement as $2:-1$.
Their parameters are totally different from ours and their spectrum has a very high $J=1$ state,
which disagrees with current experimental data.

Ref. \cite{DiPierro} studied the Dirac equation for the $\bar{Q}q$ system, but considered only the scalar confinement.
Their results again show a high $J=1$ state, while the $1s-2p$ splitting is reasonable.
But their $\alpha_s$ and $b$ seem too large.

In conclusion, it has been clarified how the masses and decay widths of the $D$ mesons behave under the change of the Lorentz properties of the confinement force.
It is interesting that the current experimental data of the decay widths prefer a large mixing of the Lorentz vector confinement.
We, however, have some difficulties in fitting all the spectrum.
We stress that the determination of the $D_1^*$ mass and width is very important to fix the picture completely.
Future experimental and theoretical studies are desirable.

\begingroup
\squeezetable
\begin{table}
\begin{tabular}{||c||c|c|c||c|c|c||}
\hline
\hline
&\multicolumn{3}{c||}{Our results ($m\!=\!300\mathrm{MeV} \>\> b\!=\!0.190\mathrm{GeV}^2$)}&\multicolumn{3}{c||}{Experimental data}\\
\hline
&$\begin{array}{c}A\\v=0\\ \frac{4\alpha_s}{3}=0.32 \end{array}$& $\begin{array}{c}B\\v=0.10 \\ \frac{4\alpha_s}{3}=0.34\end{array}$ & $\begin{array}{c}C\\v=0.40 \\ \frac{4\alpha_s}{3}=0.40\end{array}$ & Belle\cite{Belle} & CLEO\cite{CLEO} &PDG\cite{PDG}\\
\hline\hline
$M(D_2^*)\qquad[\mathrm{MeV}]$&$2459$ & $2459$ & $2459$ & $2460.7\pm 2.1 \pm 3.1$ &  &2459\\
\hline
$M(D_1^*)\qquad[\mathrm{MeV}]$&$2459$ & $2458$ & $2435$ & $2400\pm30\pm20$ & $2461^{+48}_{-42}$&\\
\hline
$M(D_1)\qquad[\mathrm{MeV}]$&$2426$ & $2421$ & $2411$ & $2423.9\pm1.7\pm0.2$ & $2422.2\pm1.8$&$2422$\\
\hline
$M(D_0^*)\qquad[\mathrm{MeV}]$&$2356$ & $2357$ & $2334$ &$2290\pm22\pm20$  &  &\\
\hline
$M(D_2^*)-M(D_1)-36.8\quad[\mathrm{MeV}]$&$-4.289$ & $1.165$ & $11.36$ &  &  &\\
\hline
$M(D_1)-M(D^*)-413.85\quad[\mathrm{MeV}]$&$-80.46$ & $-97.83$ & $-105.3$ &  & &\\
\hline
$M_{Q_c}\qquad[\mathrm{MeV}]$& $1095$ & $1351$ & $1979$ &  &  &\\
\hline
$\Gamma_{D_2^*-D(D^*)\pi}\qquad[\mathrm{MeV}]$&$67.97$ & $65.89$& $57.36$ & $46.4\pm4.4\pm3.1$ & &$25^{+8}_{-7}$\\
\hline
$\Gamma_{D_1^*-D^*\pi}\qquad[\mathrm{MeV}]$&$233.9$ & $276.2$ & $319.5$ & $380\pm100\pm100$ & $290^{+110}_{-90}$ &\\
\hline
$\Gamma_{D_1-D^*\pi}\qquad[\mathrm{MeV}]$&$75.30$ & $51.67$ & $28.06$ & $26.7\pm3.1\pm2.2$ & & $26\pm8$\\
\hline
$\Gamma_{D_0^*-D\pi}\qquad[\mathrm{MeV}]$&$360.8$ & $384.6$ & $415.0$ & $305\pm30\pm25$   &&\\
\hline
$\Gamma_{D^*-D\pi}\qquad[\mathrm{keV}]$&$194.9$ & $187.1$ & $158.1$ &  &  &$142.12\pm0.07$\\
\hline
$\begin{array}{c}\textrm{Mixing angle of}\> D_{\frac{3}{2}} \>\textrm{and}\> D_{\frac{1}{2}} \\
\> [\mathrm{degree}]\end{array}$&$13.64$ & $8.455$ & $3.178$ &  &  &\\
\hline
\hline
\end{tabular}
\caption{Our results and the experimental values of the masses and the pionic decay widths of the $D$ mesons.}
\label{ta:results}
\end{table}
\endgroup

\appendix
\section{Formulas of $D$ meson Masses}\label{ch:massformula}
We present the explicit forms of the $D$ meson masses up to $\mathcal{O}(\frac{1}{M_Q})$.
Derivatives from the momentum operator are rewritten by means of the Dirac equation (\ref{dirac}).
From the $1s_{\frac{1}{2}}$ state of the light quark, we have two states:
 \[
  |D\rangle=|J=0(1s_{\frac{1}{2}})\rangle \qquad |D^*\rangle=|J=1(1s_{\frac{1}{2}})\rangle
 \]
Their masses are given by
 \begin{eqnarray}
  \left(\!\!\begin{array}{c}
   m(D^*)\\
   m(D)
  \end{array}\!\!\right)
   \!&=&\!{m_c}^0+E_{0}+\left(\!\!\begin{array}{c}-\frac{2}{3}\\ 2\end{array}\!\!\right)\frac{1}{M_Q}\!\int \!dr{F_0(r)G_0(r)}\left(\frac{a}{r^2}+bv\right)\nonumber\\
   & & \!+\frac{a}{M_Q}\!\int\! dr \left[\left(\frac{m}{r}+b\right)\left\{{F_0}(r)\right\}^2-\left(\frac{m}{r}+b(1-2v)\right)\left\{G_0(r)\right\}^2\right]\nonumber\\
   & & \!-\frac{a}{M_Q}\!\int\! dr \left[\>\frac{1}{r}\left(\frac{a}{r}+E_0\right)\left(\left\{F_0(r)\right\}^2+\left\{G_0(r)\right\}^2\frac{}{}\right)+\frac{F_0(r)G_0(r)}{r^2}\right]\nonumber\\
   & & \!-\frac{bv}{M_Q}\!\int\! dr \left[\left(mr+br^3\right)\left\{{F_0}(r)\right\}^2-\left(mr+b(1-2v)r^3\right)\left\{G_0(r)\right\}^2\right]\nonumber\\
   & & \!+\frac{bv}{M_Q}\!\int\! dr \>r\left(\frac{a}{r}+E_0\right)\left(\left\{F_0(r)\right\}^2+\left\{G_0(r)\right\}^2\frac{}{}\right)\nonumber\\
   & & \!-\frac{1}{2M_Q}\!\int\!dr\left(\left\{m+(1-v)br\frac{}{}\!\right\}^2\!-\left\{vbr-\frac{a}{r}-E_0\right\}^2\right)\left(\left\{F_0(r)\right\}^2+\left\{G_0(r)\right\}^2\frac{}{}\!\!\right)\nonumber\\
   & & \!-\frac{1}{2M_Q}\!\int\! dr\> 2(1-v)bF_0(r)G_0(r) \label{masss1}
  \end{eqnarray}
From the $2p_{\frac{3}{2}}$ state of the light quark, we have two states:
  \[
   |D_2^*\rangle=|J=2(2p_{\frac{3}{2}})\rangle \qquad |D_{\frac{3}{2}}\rangle=|J=1(2p_{\frac{3}{2}})\rangle
  \]
Their masses are given by
  \begin{eqnarray}
   \left(\!\!\begin{array}{c}
    m(D_2^*)\\
    m(D_{\frac{3}{2}})
   \end{array}\!\!\right)
    \!&=&\!{m_c}^0+E_3+\left(\!\!\begin{array}{c}-\frac{4}{5}\\ \frac{4}{3}\end{array}\!\!\right)\frac{1}{M_Q}\!\int \!dr {F_3(r)G_3(r)}\left(\frac{a}{r^2}+bv\right)\nonumber\\
    & & \!+\frac{a}{M_Q}\!\int\! dr \left[\left(\frac{m}{r}+b\right)\left\{F_3(r)\right\}^2-\left(\frac{m}{r}+b(1-2v)\right)\left\{G_3(r)\right\}^2\right]\nonumber\\
    & & \!-\frac{a}{M_Q}\!\int\! dr \left[\>\frac{1}{r}\left(\frac{a}{r}+E_3\right)\left(\left\{F_3(r)\right\}^2+\left\{G_3(r)\right\}^2\frac{}{}\right)+\frac{2F_3(r)G_3(r)}{r^2}\right]\nonumber\\
    & & \!-\frac{bv}{M_Q}\!\int\! dr \left[\left(mr+br^3\right)\left\{{F_3}(r)\right\}^2-\left(mr+b(1-2v)r^3\right)\left\{G_3(r)\right\}^2\right]\nonumber\\
    & & \!+\frac{bv}{M_Q}\!\int\! dr \>r\left(\frac{a}{r}+E_3\right)\left(\left\{F_3(r)\right\}^2+\left\{G_3(r)\right\}^2\frac{}{}\right)\nonumber\\
    & &\!-\frac{1}{2M_Q}\int\!dr\left(\left\{m+(1-v)br\frac{}{}\!\right\}^2\!-\left\{vbr-\frac{a}{r}-E_3\right\}^2\right)\left(\left\{F_3(r)\right\}^2+\left\{G_3(r)\right\}^2\frac{}{}\!\!\right)\nonumber\\
    & &\!-\frac{1}{2M_Q}\int\! dr\> 2(1-v)bF_3(r)G_3(r) \label{massp3}
  \end{eqnarray}
From the $2p_{\frac{1}{2}}$ state of the light quark, we have two states:
 \[
  |D_{\frac{1}{2}}\rangle=|J=1(2p_{\frac{1}{2}})\rangle \qquad |D_0\rangle=|J=0(2p_{\frac{1}{2}})\rangle
 \]
Their masses are given by
 \begin{eqnarray}
  \left(\!\!\begin{array}{c}
   m(D_{\frac{1}{2}})\\
   m(D_0^*)
  \end{array}\!\!\right)
   \!&=&\!{m_c}^0+E_{1}+\left(\!\!\begin{array}{c}\frac{2}{3}\\ -2\end{array}\!\!\right)\frac{1}{M_Q}\!\int \!dr F_1(r)G_1(r)\left(\frac{a}{r^2}+bv\right)\nonumber\\
   & &\! +\frac{a}{M_Q}\!\int\! dr \left[\left(\frac{m}{r}+b\right)\left\{F_1(r)\right\}^2-\left(\frac{m}{r}+b(1-2v)\right)\left\{G_1(r)\right\}^2\right]\nonumber\\
   & &\!-\frac{a}{M_Q}\!\int\! dr \left[\>\frac{1}{r}\left(\frac{a}{r}+E_1\right)\left(\left\{F_1(r)\right\}^2+\left\{G_1(r)\right\}^2\frac{}{}\right)-\frac{F_1(r)G_1(r)}{r^2}\right]\nonumber\\
   & & \!-\frac{bv}{M_Q}\!\int\! dr \left[\left(mr+br^3\right)\left\{{F_1}(r)\right\}^2-\left(mr+b(1-2v)r^3\right)\left\{G_1(r)\right\}^2\right]\nonumber\\
   & & \!+\frac{bv}{M_Q}\!\int\! dr \>r\left(\frac{a}{r}+E_1\right)\left(\left\{F_1(r)\right\}^2+\left\{G_1(r)\right\}^2\frac{}{}\right)\nonumber\\
   & &\!-\frac{1}{2M_Q}\int\!dr\left(\left\{m+(1-v)br\frac{}{}\!\right\}^2\!-\left\{vbr-\frac{a}{r}-E_1\right\}^2\right)\left(\left\{F_1(r)\right\}^2+\left\{G_1(r)\right\}^2\frac{}{}\!\!\right)\nonumber\\
   & &\!-\frac{1}{2M_Q}\int\! dr\> 2(1-v)bF_1(r)G_1(r) \label{massp1}
 \end{eqnarray}
The radial wave functions for the $1s_{\frac{1}{2}}$, $2p_{\frac{1}{2}}$ and $2p_{\frac{3}{2}}$ states are labeled by $0$, $1$ and $3$, respectively.
The off-diagonal element of the mixing matrix of $D_{\frac{1}{2}}$ and $D_{\frac{3}{2}}$ is given by
\begin{equation}
  m_{\mathrm{mix}}=-\frac{\sqrt{2}}{3}\frac{1}{M_Q}\!\int\!dr \left({F_3(r)G_1(r)+F_1(r)G_3(r)}\frac{}{}\!\right)\left(\frac{a}{r^2}+bv\right) \label{mixoff}
 \end{equation}
Then the eigenvalues are determined by the secular equation given by   
\begin{equation}
  \left|\begin{array}{cc}
   m(D_{\frac{1}{2}})-\lambda & m_{\mathrm{mix}} \\
   m_{\mathrm{mix}} & m(D_{\frac{3}{2}})-\lambda 
  \end{array}\right|=0 \label{mixmat}
 \end{equation}
Solving this equation, the masses of $D_1^*$ and $D_1$ are given by 
 \begin{eqnarray}
  m(D_1^*)&=&\frac{m(D_{\frac{1}{2}})+m(D_{\frac{3}{2}})+\sqrt{\left(m(D_{\frac{1}{2}})-m(D_{\frac{3}{2}})\right)^2+{m_{\mathrm{mix}}}^2}}{2}\nonumber\\
  m(D_1)&=&\frac{m_(D_{\frac{1}{2}})+m(D_{\frac{3}{2}})-\sqrt{\left(m(D_{\frac{1}{2}})-m(D_{\frac{3}{2}})\right)^2+{m_{\mathrm{mix}}}^2}}{2} \label{d1d1star}
 \end{eqnarray} 
It is conventional to assign $D_1^*$ to the state with larger $D_{\frac{1}{2}}$ component.
In our case, $D_1^*$ happens always to be heavier than $D_1$. 

\end{document}